\documentclass[journal]{IEEEtran}

\usepackage{amsmath,bm,amssymb,amsfonts,amsthm}
\usepackage{algorithm,algorithmic}
\usepackage{graphicx}
\usepackage{xcolor}
\usepackage{relsize}
\usepackage{textcomp}
\usepackage{listings}

\usepackage{changepage}
\usepackage{hhline}
\usepackage{multirow}
\usepackage{nomencl}
\usepackage{mathtools}
\usepackage{commath}
\usepackage{booktabs}
\usepackage{subfiles}
\usepackage{tabularx}
\usepackage{lscape}
\usepackage{rotating}

\usepackage[normalem]{ulem}
\usepackage[utf8]{inputenc}
\usepackage[hyphens]{url}
\usepackage{lineno}
    \modulolinenumbers[5]
\usepackage[caption=false,font=footnotesize]{subfig}
\usepackage{float}
\usepackage{enumitem}
    \setlist{nolistsep}
\usepackage{scalerel}
\usepackage{tikz}
\usetikzlibrary{svg.path}
\definecolor{orcidlogocol}{HTML}{A6CE39}
\tikzset{
  orcidlogo/.pic={
    \fill[orcidlogocol] svg{M256,128c0,70.7-57.3,128-128,128C57.3,256,0,198.7,0,128C0,57.3,57.3,0,128,0C198.7,0,256,57.3,256,128z};
    \fill[white] svg{M86.3,186.2H70.9V79.1h15.4v48.4V186.2z}
                 svg{M108.9,79.1h41.6c39.6,0,57,28.3,57,53.6c0,27.5-21.5,53.6-56.8,53.6h-41.8V79.1z M124.3,172.4h24.5c34.9,0,42.9-26.5,42.9-39.7c0-21.5-13.7-39.7-43.7-39.7h-23.7V172.4z}
                 svg{M88.7,56.8c0,5.5-4.5,10.1-10.1,10.1c-5.6,0-10.1-4.6-10.1-10.1c0-5.6,4.5-10.1,10.1-10.1C84.2,46.7,88.7,51.3,88.7,56.8z};
  }
}
\newcommand\orcidicon[1]{\href{https://orcid.org/#1}{\mbox{\scalerel*{
\begin{tikzpicture}[yscale=-1,transform shape]
\pic{orcidlogo};
\end{tikzpicture}
}{|}}}}
\usepackage[bookmarksopen,bookmarksdepth=3]{hyperref}

\usepackage[noadjust]{cite}

\begin{document}

\title{\huge Implementing Admittance Relaying for Microgrid Protection}

\author{
    Arthur~K.~Barnes $^{1}$\orcidicon{0000-0001-9718-3197} \IEEEmembership{Member, IEEE}
    and~Adam~Mate $^{1}$\orcidicon{0000-0002-5628-6509} \IEEEmembership{Member, IEEE}

\thanks{Manuscript submitted: December~14,~2020. 
}

\thanks{$^{1}$ The authors are with the Advanced Network Science Initiative at Los Alamos National Laboratory, Los Alamos, NM 87544 USA. Email:\{abarnes, amate\}@lanl.gov.}

\thanks{Color versions of one or more of the figures in this paper are available online at https://ieeexplore.ieee.org.}

\thanks{LANL ANSI LA-UR-20-30128.}

}

\markboth{IEEE/IAS 57th Industrial \& Commercial Power Systems Technical Conference, April~2021}{}

\maketitle

\begin{abstract}
The rapid increase of distributed energy resources has led to the widespread deployment of microgrids. These flexible and efficient local energy grids are able to operate in both grid-connected mode and islanded mode; they are interfaced to the main power system by a fast semiconductor switch and commonly make use of inverter-interfaced generation. 
This paper focuses on inverter interfaced microgrids, which present a challenge for protection as they do not provide the high short-circuit current necessary for conventional time-overcurrent protection. The application of admittance relaying for the protection of inverter-interfaced microgrids is investigated as a potential solution. The comparison of analytical and simulated results of performed four experiments prove the suitability of admittance relaying for microgrids protection. 
\end{abstract}


\section{Introduction} \label{sec:intro}
\indent

In their recent paper, McDermott et al.~\cite{mcdermott_protective_2018} evaluated issues with microgrid protection. They highlighted the underlying difficulties, which include the lack of fault current from inverter-interfaced generation \cite{tumilty_approaches_2006}, the varying fault current between grid-connected and islanded modes \cite{tumilty_approaches_2006}, the potential for normally-meshed operation \cite{dewadasa_line_2008} and unbalanced operation due to single-phase loads \cite{dewadasa_line_2008}.
A handful of possible solutions (with some severe limitations) have been proposed for these challenges over the years.
Dewadasa et al. \cite{dewadasa_line_2008, dewadasa_distance_2008} identified admittance protection for load protection; however, there is a potential for loss of protection selectivity in the case of upstream line-ground faults \cite{mcdermott_protective_2018}.
Kar et al.~\cite{kar_time-frequency_2014} identified differential protection based on the discrete S-transform for line protection; however, there is a potential for blinding the protection when fault contribution on either end of the line is similar \cite{mcdermott_protective_2018}.
Barnes et al.~\cite{barnes21-dse} identified dynamic state estimation for the protection of radial portions of microgrids; however, there is a potential for sensitivity to initial conditions, particularly at the delta-connected load models.
Nevertheless, it is worth noting that not all microgrid designs present these above discussed challenges; for example, microgrids could choose to omit meshed operation.

This paper investigates the application of admittance relaying for the protection of inverter-interfaced microgrids.
Section~\ref{sec:background} describes the relevant background of admittance and pilot protection as solutions for line and load protection, in addition to describing the protection schemes analyzed and the behavior of inverter-interfaced microgrids under faults. Section~\ref{sec:seq-networks} derives sequence networks for the fault cases considered.
Section~\ref{sec:transient-model} presents the transient model constructed to validate the impedances calculated from the sequence networks.
Section~\ref{sec:results} compares analytical and simulated results from the sequence networks and transient model.
Finally, Section~\ref{sec:conclusions} summarizes the conclusions of this paper in terms of the suitability of admittance relaying for microgrid protection.

\section{Background} \label{sec:background}

\subsection{Admittance Protection}

For protection, the same quantities are used as that of Dewadasa \cite{dewadasa_line_2008, dewadasa_distance_2008, blackburn_protective_2007}, where line-ground faults are detected by:
\begin{equation*}
Z_{lg}^1 = \frac{V_{ar}}{I_{ar} + k I_{ar}^0} 
\end{equation*}
\noindent and line-line faults are detected by:
\begin{equation*}
Z_{ll}^1 = \frac{V_{cr} - V_{br}}{I_{br} - I_{cr}}
\end{equation*}
\noindent where $Z_{lg}^1$ and $Z_{ll}^1$ are the measured positive-sequence impedances (measured by the phase and ground distance relaying); $V_{ar}$, $V_{br}$, and $V_{cr}$ are the measured phase-ground voltages; $I_{ar}$, $I_{br}$, and $I_{cr}$ are the measured phase currents; and $k$ is related to the ratio of positive- and negative-sequence line impedance:
\begin{equation*}
k = 1 - \frac{Z^0}{Z^1}
\end{equation*}

Both $Z_{lg}^1$ and $Z_{ll}^1$ measure the positive-sequence impedance between the relay and the fault, which allows them to accurately measure the distance to the fault for practical lines of sufficient length with mutual impedance between conductors.

Admittance protection on its own is suitable for load bus protection.
For line protection, as line lengths are low in microgrids, it may be necessary to use pilot protection. This means that the likelihood of protection erroneously determining if a fault is in- or out-of-zone is high \cite{blackburn_protective_2007}.

\subsection{Pilot Protection}

\begin{figure}[!htbp]
\centering
\includegraphics[scale=0.6]{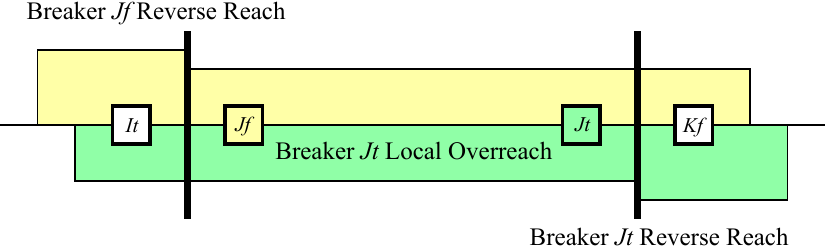}
\caption{Oneline diagram for a DCB scheme.}
\label{fig:dcb-oneline}
\end{figure}

\begin{figure}[!htbp]
\centering
\includegraphics[scale=0.265]{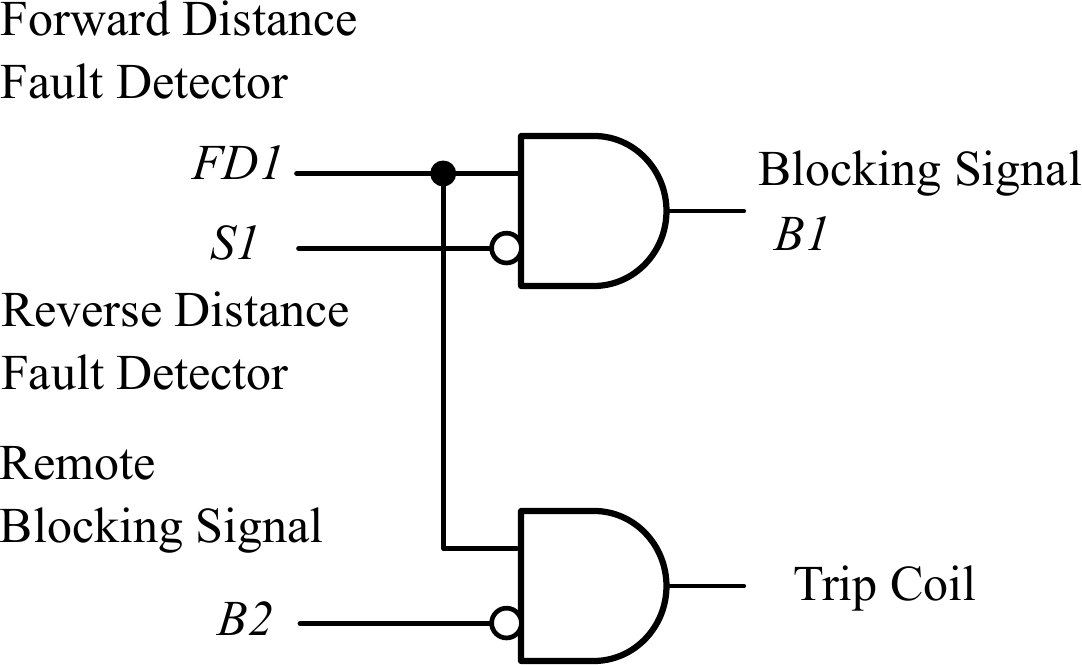}
\caption{Schematic for one relay in a DCB scheme.}
\label{fig:dcb-logic}
\end{figure}

The recommended pilot protection scheme is directional comparison blocking (abbr.~DCB).
Typically in this scheme, a line is protected by two directional distance relays on either end for line-line faults and directional overcurrent relays for line-ground faults \cite{blackburn_protective_2007, tamronglak_analysis_1994, elizondo_methodology_2003}.
A schematic for one such relay in oneline diagram from  Fig.~\ref{fig:dcb-oneline} is illustrated in Fig.~\ref{fig:dcb-logic}.
For inverter-interfaced microgrid protection, as it is not possible to rely on the presence of fault currents, it is necessary to use ground distance relaying instead of overcurrent relaying \cite{tumilty_approaches_2006}.

DCB operates with a bidirectional channel along the line by which each relay can send a blocking signal preventing the other from operating. An important feature of this scheme is that the channel is only transmitting if a relay detects a fault and is actively sending the blocking signal. The scheme is therefore biased towards dependability in that the breakers on either end of the line can operate if the communication channel is inoperable.
If a relay's directional element detects that the fault is external, it will send a blocking signal to the relay on the other end of the line to prevent it from operating. DCB is therefore suitable for transmission line taps or radial portions of a microgrid where the relay on the receiving end of the line may not experience fault current, causing an unblocking scheme could fail to operate.

\subsection{Inverter-Interfaced Microgrids}

Generation with inverter front-ends in microgrids is expected to supply single-phase loads in addition to balanced three-phase loads.
Proportional-resonant (abbr.~PR) controllers offer better performance during unbalanced operation and in the presence of load harmonics compared to proportional-integral (abbr.~PI) controllers in a rotating reference frame (e.g., DQ0) \cite{teodorescu_pr_2003,vasquez_modeling_2013}.
PR controllers can operate in a stationary reference frame that can either be the raw ``\textit{A, B, C}'' coordinates or the ``$\alpha$, $\beta$'' coordinates produced by the Clarke transformation \cite{teodorescu_pr_2003}. In case the inverter supplies a four-wire system, an additional controller for the ``$\gamma$'' coordinate, corresponding to zero-sequence quantities, is necessary \cite{bottrell_comparison_2014}; therefore it may be preferable to remain in ``\textit{A, B, C}'' coordinates.
In this paper, a three-phase inverter with no neutral connection is selected and a delta-grounded wye transformer provides a source for zero-sequence current.

A major distinction between inverter-based generation and rotating machinery is that the thermal time-constants for power electronic semiconductor modules are very low, consequently inverters are unable to supply fault current for the amount of time it takes for conventional protection to clear a fault.
Two main options are available for protecting inverters from overheating during faults: instantaneous saturation and latching current limiters \cite{bottrell_comparison_2014}.
In an inverter with an inner current loop and outer voltage or power loop, such as that described in \cite{vasquez_modeling_2013}, the current limiters operate on the output of the voltage loop, which acts as a reference to the current loop.
In instantaneous saturation, the current reference is simply bounded within an allowable range; this introduces considerable voltage harmonics into the system, which can complicate protection.
Latched current limiters avoid harmonic injection during faults, but can introduce a current discontinuity when switching the current controller reference from the voltage controller output to the limited current signal \cite{bottrell_comparison_2014}.

\section{Equivalent Sequence Networks} \label{sec:seq-networks}
\indent

\begin{figure*}[!htbp]
\centering
\includegraphics[scale=0.375]{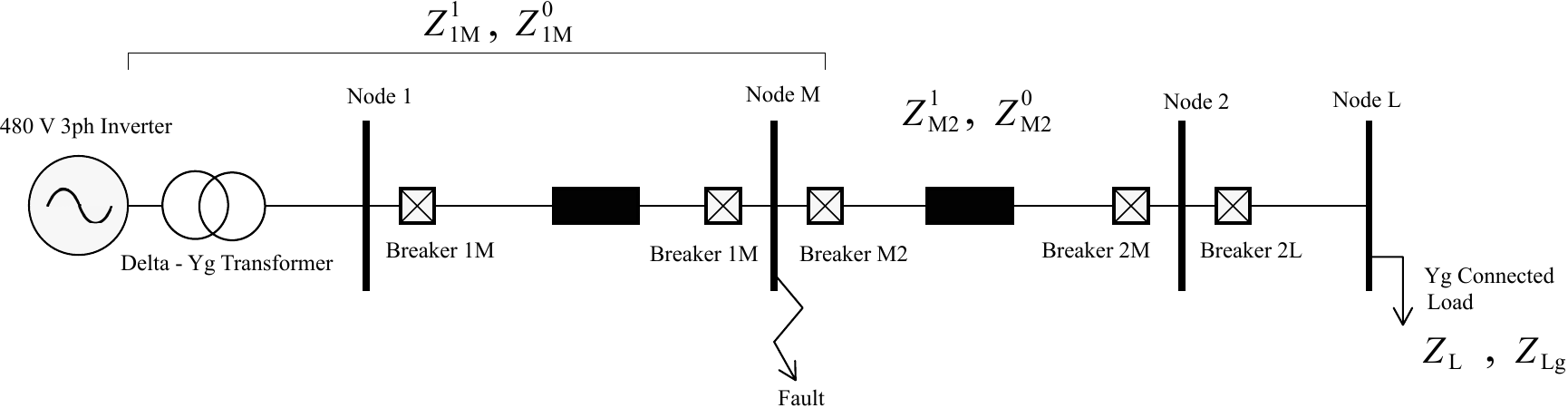}
\caption{Oneline diagram for the considered two bus microgrid system.}
\label{fig:ug-oneline}
\end{figure*}

The system considered is a two bus microgrid, illustrated in Fig.~\ref{fig:ug-oneline}; the system parameters are presented later in Section~\ref{sec:transient-model}.

\noindent Six cases are considered for deriving sequence networks:
\begin{enumerate}
\item Line-ground fault for an ideal voltage source with an upstream relay.
\item Line-ground fault for a current-limiting inverter with an upstream relay.
\item Line-ground fault with a downstream relay.
\item Line-line fault for an ideal voltage source with an upstream relay.
\item Line-line fault for a current-limiting inverter with an upstream relay.
\item Line-line fault with a downstream relay. 
\end{enumerate}
\vspace{0.1in}

It is shown that the impedance calculations for downstream relays are not affected by choice of an ideal voltage source or a current-limiting inverter.

\subsection{Line-Ground Midpoint Faults}

\noindent \textit{1) Ideal Voltage Source and Upstream Relay}

\begin{figure*}[!htbp]
\centering
\includegraphics[scale=0.35,angle=0]{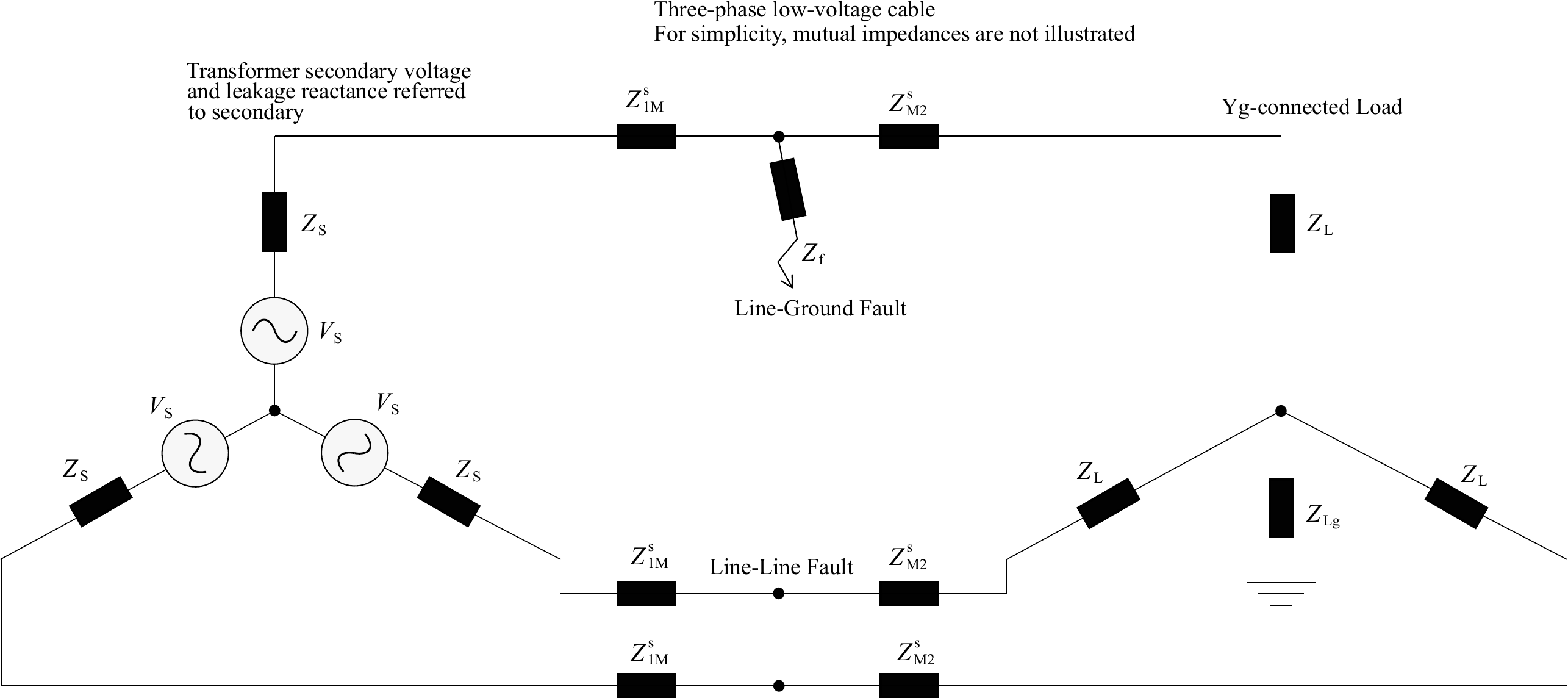}
\caption{Thevenin equivalent circuit of the microgrid.}
\label{fig:ug-3ph}
\end{figure*}

To analyze the behavior of the considered microgrid during faults, it is necessary to create the Thevenin equivalent circuit of the system from the perspective of the protective relaying.
The first step is to draw out the relevant portions with all conductors depicted, as seen in Fig.~\ref{fig:ug-3ph}; note that the microgrid inverter and the delta winding of the transformer are not illustrated.
As mentioned earlier, the microgrid uses an inverter with a three-phase H-bridge connected to a delta-wye transformer with a grounded wye. Rather than using three H-bridges to provide neutral currents (as in \cite{dewadasa_line_2008}), this system relies on the transformer grounding to do so.

\begin{figure}[!htbp]
\centering
\subfloat[Equivalent positive-sequence network]{\includegraphics[scale=0.375]{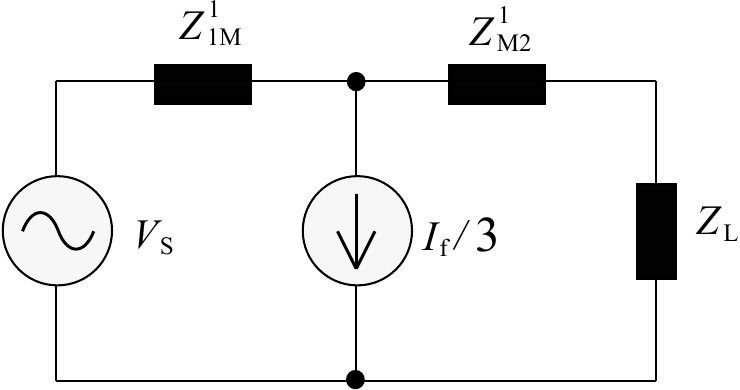}}

\subfloat[Equivalent negative-sequence network]{\includegraphics[scale=0.375]{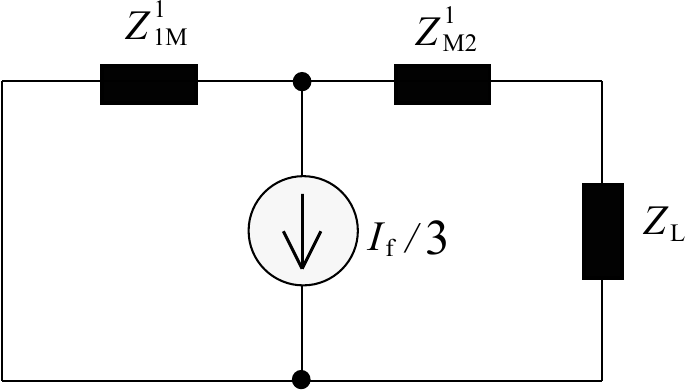}}

\subfloat[Equivalent zero-sequence network]{\includegraphics[scale=0.375]{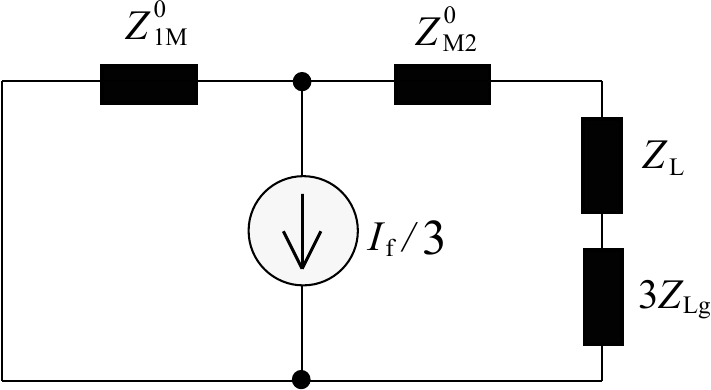}}

\vspace{0.1in}
\caption{Sequence networks for a midpoint line-ground fault with an ideal voltage source.}
\label{fig:ug-fault-lg-midpoint-seqs}
\end{figure}

Fig.~\ref{fig:ug-fault-lg-midpoint-seqs} illustrates the equivalent sequence networks of the microgrid for the case of a midpoint line-ground fault and ideal voltage source.
The current flowing into the fault is identical for each circuit: $I_f^1 = I_f^2 = I_f^0 = I_f/3$. This means that the sequence equivalent circuits can be linked in series.

For the positive-sequence network, the Thevenin equivalent impedance:
\begin{equation*}
Z_{eq1} = Z_{1M}^1 || (Z_{M2}^1 + Z_L)
\end{equation*}
\noindent where $||$ denotes the equivalent impedance of parallel circuit elements:
\begin{equation*}
Z_x || Z_y = \frac{Z_xZ_y}{Z_x + Z_y}
\end{equation*}
\noindent The Thevenin equivalent voltage:
\begin{equation*}
V_{eq1} = V_S \cdot \frac{Z_{M2}^1 + Z_L}{Z_{1M}^1 + (Z_{M2}^1 + Z_L)}
\end{equation*}

\noindent For the negative-sequence network, the Thevenin equivalent impedance is the same as $Z_{eq1}$:
\begin{equation*}
Z_{eq2} = Z_{eq1} = Z_{1M}^1 || (Z_{M2}^1 + Z_L).
\end{equation*}

\noindent Last, for the zero-sequence network, the Thevenin equivalent impedance:
\begin{equation*}
Z_{eq0} = Z_{1M}^0 || (Z_{M2}^0 + Z_L + 3Z_{Lg}).
\end{equation*}

\begin{figure}[!htbp]
\centering
\includegraphics[scale=0.375]{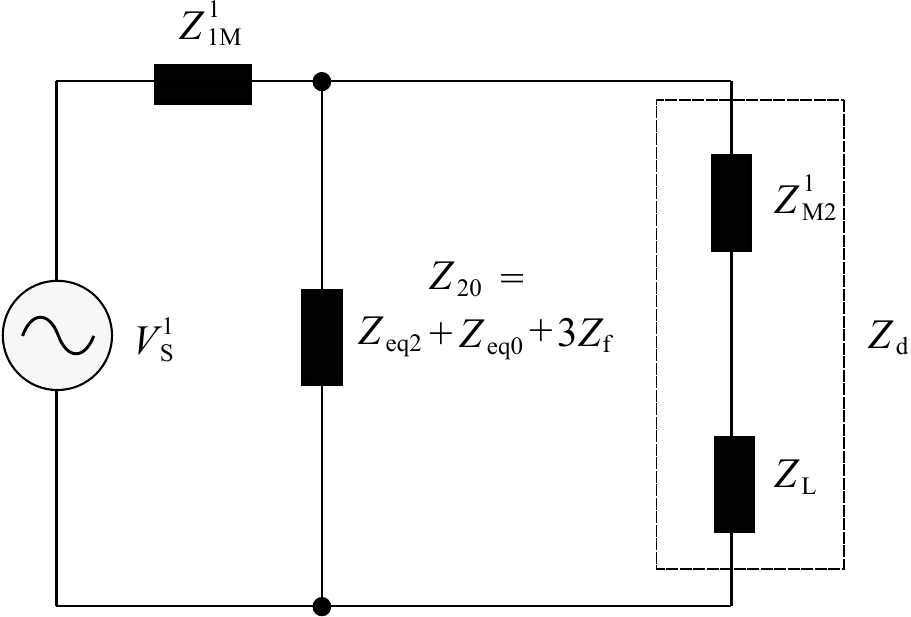}
\caption{Simplified sequence network for a midpoint line-ground fault with an ideal voltage source.}
\label{fig:ug-fault-lg-midpoint-012-bal-simpl}
\end{figure}

This set of interconnected sequence networks can be simplified, as illustrated in Fig.~\ref{fig:ug-fault-lg-midpoint-012-bal-simpl}.

Given a protective relay at the load bus, the current flowing though the relay can be determined by calculating the current flowing through the relay in the positive-, negative-, and zero-sequence networks.

\noindent First, the impedance of the positive-sequence network downstream of the relay:
\begin{equation*}
Z_d = Z_{M2}^1 + Z_L
\end{equation*}

\noindent Next, the equivalent impedance of the negative-sequence network, zero-sequence network, and fault:
\begin{equation*}
Z_{20} = Z_{eq2} + Z_{eq0} + 3Z_f
\end{equation*}
\noindent therefore the equivalent impedance downstream of the relay:
\begin{equation*}
Z_{20d} = Z_{20} || Z_d
\end{equation*}

\noindent The measured positive-sequence current at the relay:
\begin{equation*}
I_{r}^1 = \frac{V_s^1}{Z_{1M} + Z_{20d}}
\end{equation*}

\noindent To calculate the negative- and zero-sequence currents through the relay, the first step is to calculate the Norton equivalent of the voltage source:
\begin{equation*}
I_{sn} = \frac{V_s^1}{Z_{1M}^1}
\end{equation*}

\begin{figure}[!htbp]
\centering
\includegraphics[scale=0.375]{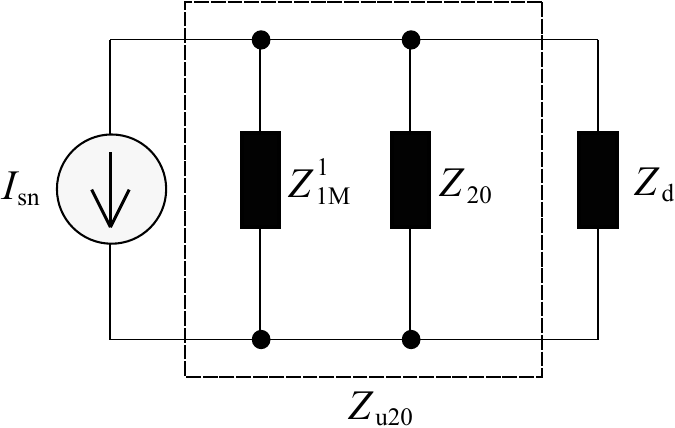}
\caption{Norton equivalent circuit for a midpoint line-ground fault with an ideal voltage source.}
\label{fig:ug-fault-lg-midpoint-bal-simpl-norton}
\end{figure}

\noindent The equivalent impedance of the upstream and downstream portions of the positive-sequence network in parallel:
\begin{equation*}
Z_{1d} = Z_{1M}^1 || Z_{d}
\end{equation*}
\noindent which produces the simplified Norton equivalent circuit, illustrated in Fig.~\ref{fig:ug-fault-lg-midpoint-bal-simpl-norton}.

Using a current divider and assuming that $Z_{1M}^1 \ll Z_L$, the negative- and zero-sequence current flowing through the relay:
\begin{equation*}
I_{r}^2 = I_{r}^0 = I_{sn}\frac{Z_{1d}}{Z_{20} + Z_{1d}}
\end{equation*}
\noindent therefore the measured phase A current at the relay:
\begin{equation*}
I_{r}^a = I_{r}^1 + I_{r}^2 + I_{r}^0
\end{equation*}  
\noindent and the voltage at the relay:
\begin{equation*}
V_r^a = V_s^1 - Z_{1M}I_r^1 + Z_{1M}^1 I_r^2 + Z_{1M}^0 I_r^0
\end{equation*}
\noindent finally, the positive-sequence impedance can be calculated as:
\begin{equation*}
Z_r^1 = \frac{V_r^{ag}}{I_r^a}
\end{equation*}

\begin{figure}[!htbp]
\centering
\includegraphics[scale=0.5]{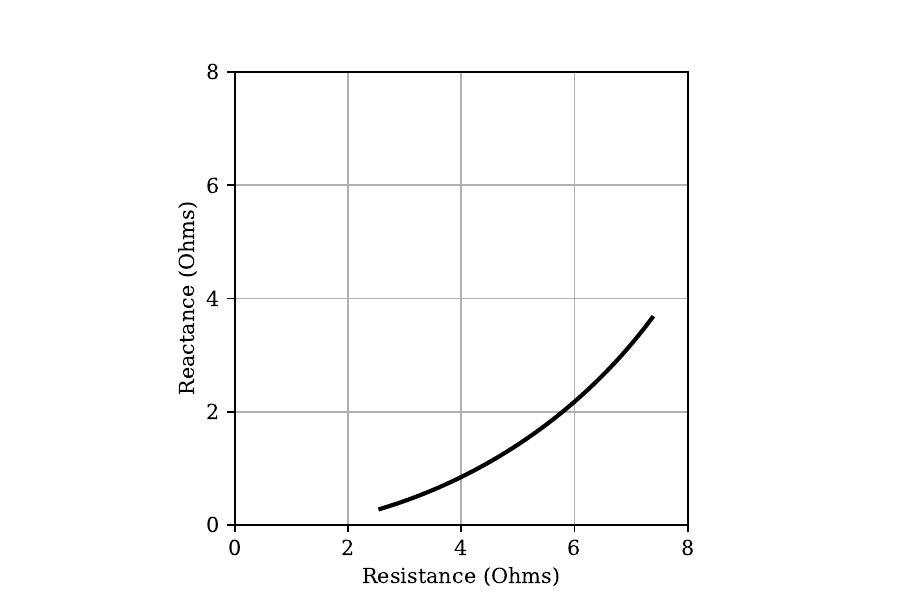}
\caption{Measured positive-sequence impedance at an upstream relay for a midpoint line-ground fault with an ideal voltage source.}
\label{fig:zr1-ug-fault-lg-midpoint-balanced}
\end{figure}

\noindent The measured positive-sequence impedance at the relay, as fault current varies, is illustrated in Fig.~\ref{fig:zr1-ug-fault-lg-midpoint-balanced} (the fault resistance is varied from 3.68~$[\Omega$] to 1~[$k\Omega$] and the lower bound is selected for a fault current twice that of the load current).

\newpage
\noindent \textit{2) Current-Limiting Inverter and Upstream Relay}

\begin{figure}[!htbp]
\centering
\includegraphics[scale=0.375]{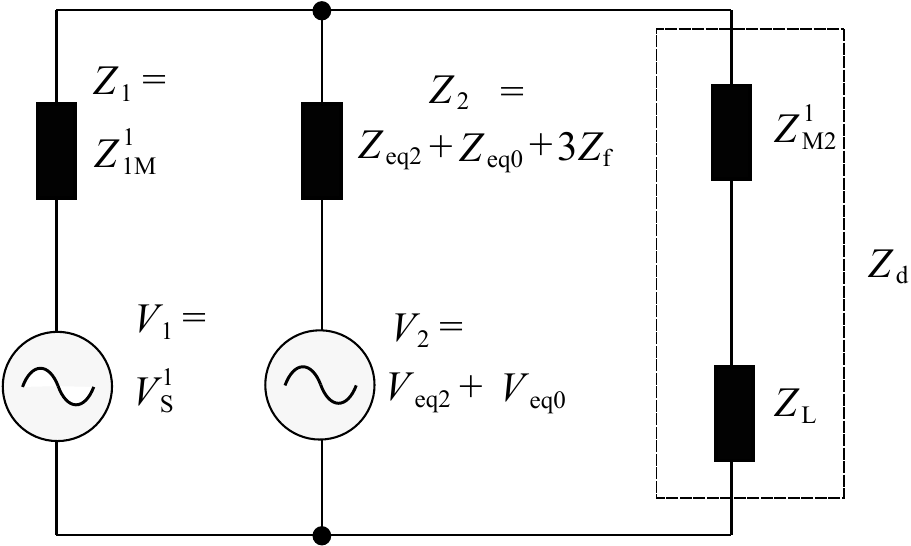}
\caption{Simplified sequence network for a midpoint line-ground fault with a current-limiting inverter.}
\label{fig:ug-fault-lg-midpoint-012-simpl}
\end{figure}

The current-limiting inverter is modeled as an unbalanced voltage source for the simplicity of calculation (as opposed to voltage sources on the unfaulted phases and current sources on the faulted phases) \cite{dewadasa_line_2008}. This results in voltage sources present on each of the sequence networks.
The values of the inverter negative- and zero-sequence voltages are determined by simulation (described in Section \ref{sec:transient-model}) and are calculated to be roughly 60\% of the positive-sequence voltage.
The simplified sequence network is illustrated in Fig.~\ref{fig:ug-fault-lg-midpoint-012-simpl}.

The current through the relay can be calculated given:
\begin{align*}
V_1 = & V_s^1 &
Z_1 = & Z_{1M}^1 \\
V_2 = & V_{eq2} + V_{eq0} &
Z_2 = & Z_{eq2} + Z_{eq0} + 3Z_f \\ & &
Z_d = & Z_{M2}^1 + Z_L
\end{align*}

\noindent Using parallel branch combinations:
\begin{align*}
Z_{1d} & = Z_1 || Z_d &
Z_{2d} & = Z_2 || Z_d
\end{align*}
\noindent and Norton equivalent sources:
\begin{align*}
I_{1n} & = \frac{V_1}{Z_1} &
I_{2n} & = \frac{V_2}{Z_2}
\end{align*}
\noindent the positive- and negative-sequence currents through the relay are calculated as follows:

First, the negative- and zero-sequence voltage sources are disabled. The positive-sequence current, resulting from the positive-sequence voltage source, measured flowing through the relay:
\begin{equation*}
I_{11} = \frac{V_1}{Z_1 + Z_{2d}}
\end{equation*}

\begin{figure}[!htbp]
\centering
\includegraphics[scale=0.375]{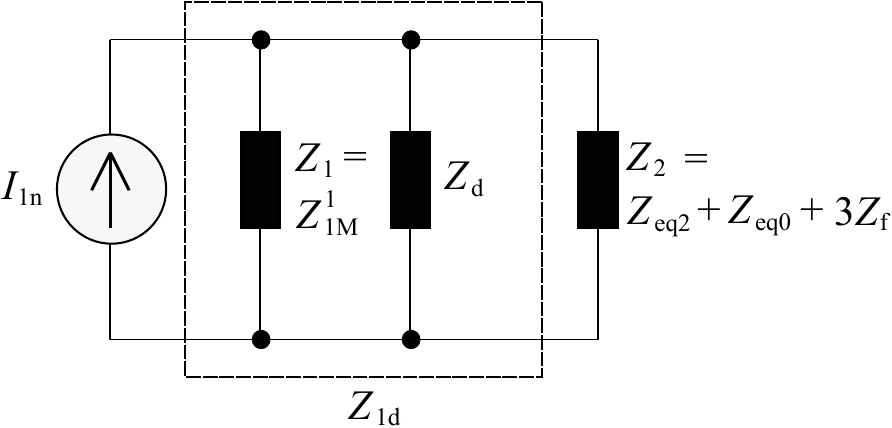}
\caption{Norton equivalent circuit for a line-ground fault, supplied by a current-limiting inverter, with the negative-sequence and zero-sequence voltage sources disabled.}
\label{fig:ug-fault-lg-midpoint-012-norton-v1}
\end{figure}

Second, the Norton equivalent circuit (illustrated in Fig.~\ref{fig:ug-fault-lg-midpoint-012-norton-v1}), with the negative- and zero-sequence voltage sources disabled, is used. The negative-sequence current, resulting from the positive-sequence voltage source, flowing through the relay:
\begin{equation*}
I_{21} \approx I_{1n}\frac{Z_{1d}}{Z_2 + Z_{1d}}
\end{equation*}
\noindent assuming that $Z_{1M}^1 \ll Z_d$

\begin{figure}[!htbp]
\centering
\includegraphics[scale=0.375]{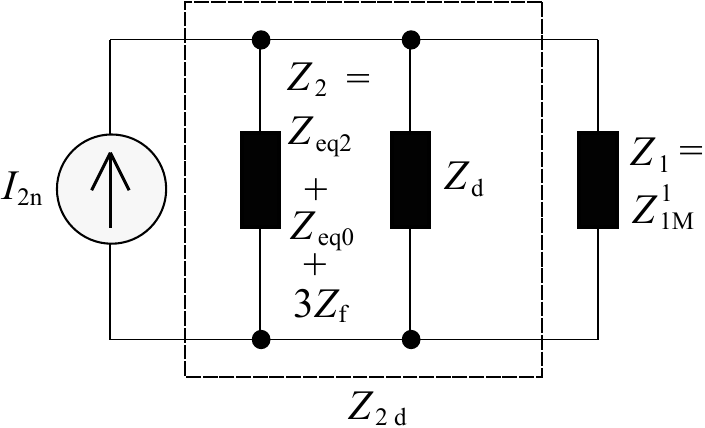}
\caption{Norton equivalent circuit for a line-ground fault, supplied by a current-limiting inverter, with positive-sequence voltage source disabled.}
\label{fig:ug-fault-lg-midpoint-012-norton-v2}
\end{figure}

Third, the Norton equivalent circuit (illustrated in Fig.~\ref{fig:ug-fault-lg-midpoint-012-norton-v2}), with the positive-sequence voltage source disabled, is used. The positive-sequence current, from the contributions of the negative- and zero-sequence voltage sources, flowing through the relay:
\begin{equation*}
I_{12} = I_{2n}\frac{Z_{2d}}{Z_1 + Z_{2d}}
\end{equation*}

Finally, the negative-sequence current contribution from the negative- and zero-sequence sources can be calculated:
\begin{equation*}
I_{22} \approx \frac{V_2}{Z_2 + Z_{1d}}
\end{equation*}
\noindent again assuming that $Z_{1M}^1 \ll Z_L$

\noindent The measured current running through the relay:
\begin{align*}
I_r^1 & = I_{11} + I_{12} \\
I_r^2 & = I_{21} + I_{22} \\
I_r^0 & = I_r^2 \\
I_r^a & = I_r^1 + I_r^2 + I_r^0
\end{align*}

\noindent The measured positive-sequence impedance:
\begin{equation*}
Z_r = \frac{V_r^{ag}}{I_r^a}
\end{equation*}

The measured positive-sequence impedance at the relay, as a function of the fault resistance, is illustrated in Fig.~\ref{fig:zr1-ug-fault-lg-midpoint-balanced}.

\begin{figure}[!htbp]
\centering
\includegraphics[scale=0.5]{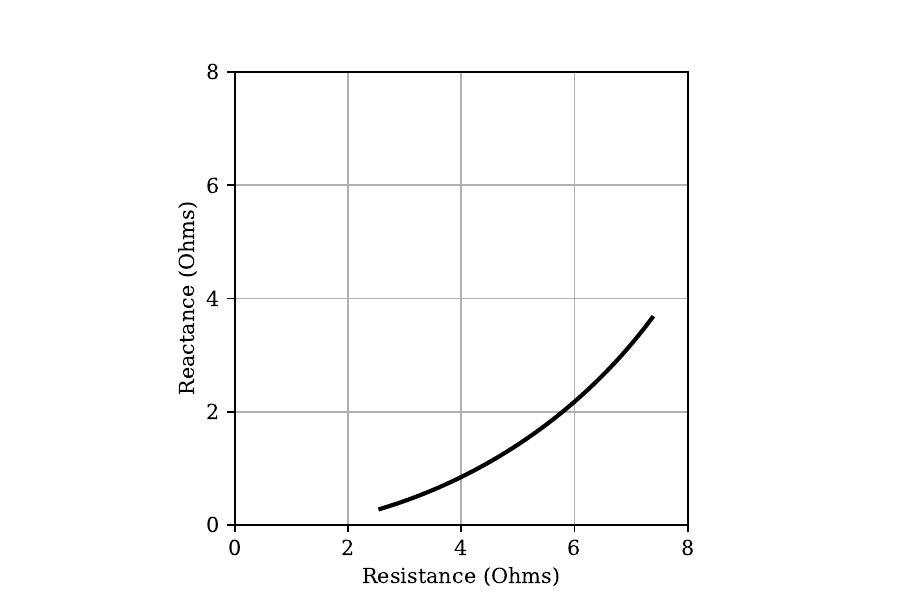}
\caption{Measured positive-sequence impedance at an upstream relay for a midpoint line-ground fault with a current-limiting inverter.}
\label{fig:zr1-ug-fault-lg-midpoint}
\end{figure}

\newpage
\noindent \textit{3) Downstream Relay}

By using the definition of sequence voltages \cite{bergen_power_2009} and Fig.~\ref{fig:ug-fault-lg-midpoint-012-simpl}, the measured voltage across the relay:
\begin{equation*}
V_r^{ag} = Z_d^1 I_r^1 + Z_d^1 I_r^1 + Z_d^0 I_r^0
\end{equation*}
\noindent where
\begin{align*}
Z_d^1 = & Z_{M2}^1 + Z_L &
Z_d^0 = & Z_{M2}^0 + Z_L + 3Z_{Lg}
\end{align*}

\noindent The measured impedance by the relay:
\begin{align*}
Z_r^1 &= \frac{V_r^{ag}}{I_r^a + kI_r^0} \\
&= \frac{Z_d^1[I_r^1 + I_r^2 + (Z_d^0/Z_d^1)I_r^0]}{I_r^a + kI_r^0}
\end{align*}

\noindent It is therefore apparent that $Z_r = Z_d^1$ and relay is not going to operate for line-ground faults when $k = 1 - Z_d^0/Z_d^1$, unless it experiences load encroachment.

\subsection{Line-Line Midpoint Faults}

For line-line midpoint faults, to facilitate expressing the fault quantities analytically, the analysis assumes that the fault impedance is small (but non-zero).
\vspace{0.05in}

\vspace{0.1in}
\noindent \textit{1) Ideal Voltage Source and Upstream Relay}

\begin{figure}[!htbp]
\centering
\includegraphics[scale=0.33]{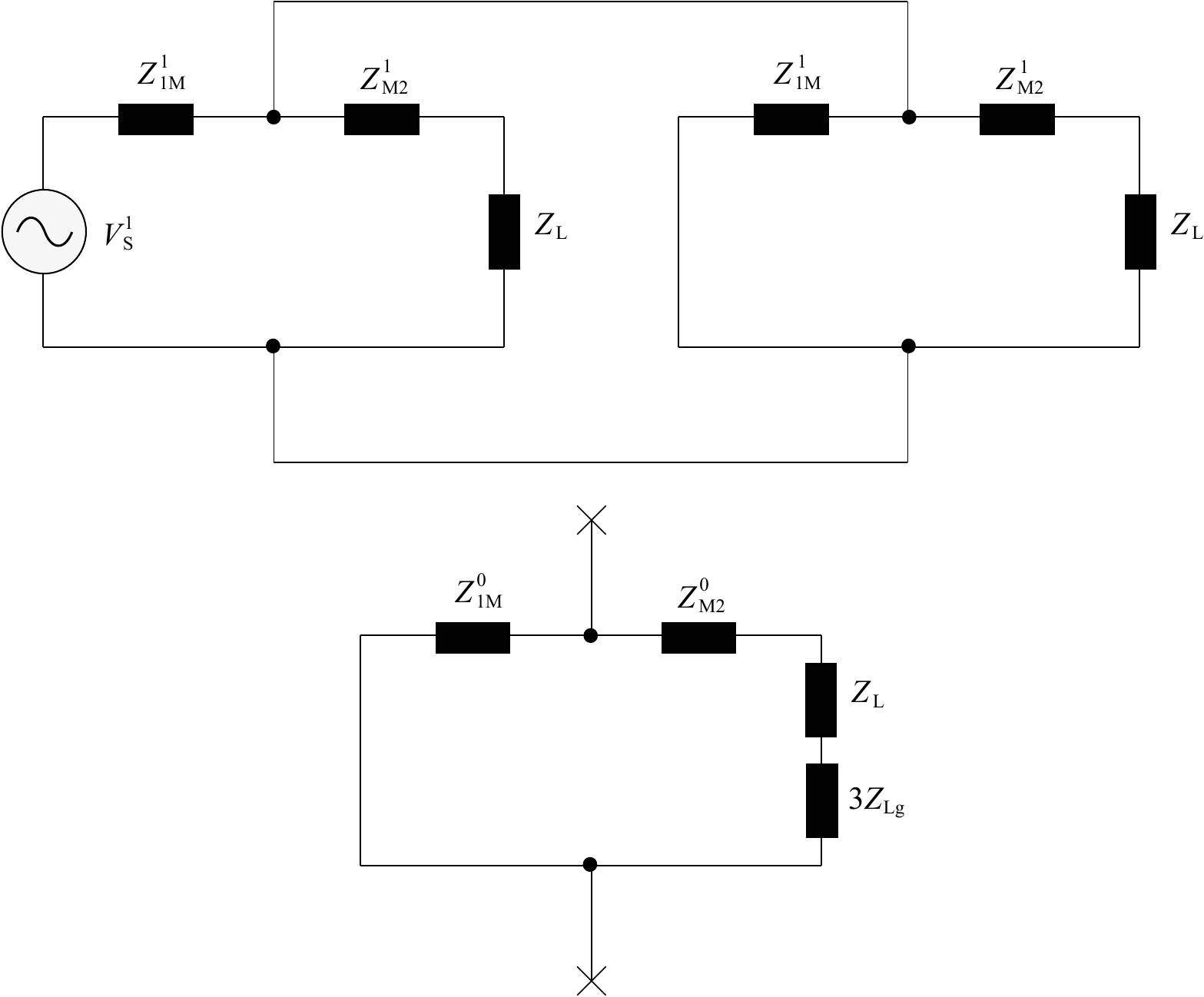}
\caption{Sequence networks for a midpoint line-line fault with an ideal voltage source.}
\label{fig:ug-fault-ll-midpoint-012-bal}
\end{figure}

Fig.~\ref{fig:ug-fault-ll-midpoint-012-bal} illustrates the equivalent sequence networks of the microgrid for the case of a midpoint line-line fault across phases B and C.
It is apparent from the figure that the currents flowing into the fault are equal in magnitude, but with opposite signs: $I_{bf}$ = $-I_{cf}$. Furthermore, the line-ground voltages at the fault location on phases B and C are equal: $V_{bf}$ = $V_{cf}$.

From the definition of sequence quantities, it can be shown that $V_{af}^1$ = $V_{af}^2$ and $I_{af}^1$ = $-I_{af}^2$. This implies that the positive- and negative-sequence networks are connected in parallel via the fault, while the zero-sequence network is isolated from the other two (as shown in Fig.~\ref{fig:ug-fault-ll-midpoint-012-bal}).
Given that there is no source of zero-sequence voltage, there is no zero-sequence current flowing through the system for this fault type.

\begin{figure}[!htbp]
\centering
\includegraphics[scale=0.375]{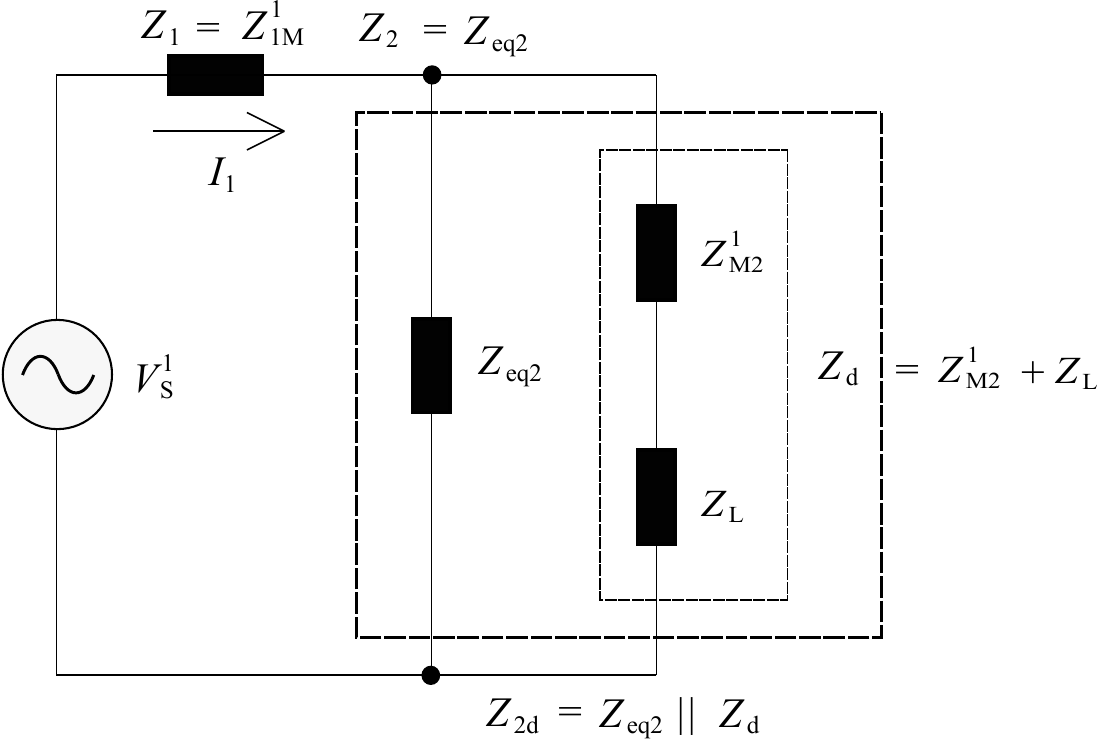}
\caption{Simplified sequence network for a midpoint line-line fault with an ideal voltage source.}
\label{fig:ug-fault-ll-midpoint-012-bal-simpl}
\end{figure}

\newpage
The connected positive- and negative-sequence networks can be simplified, as illustrated in Fig.~\ref{fig:ug-fault-ll-midpoint-012-bal-simpl}.
As is the case for the line-ground fault, the upstream impedance is $Z_1$ = $Z_{1M}^1$, the downstream impedance is $Z_d$ = $Z_{M2}^1$ + $Z_L$, and the shunt impedance from the negative-sequence network is $Z_2$ = $Z_{eq2}$, where $Z_{eq2}$ is the same as defined in Section~III-A-1.

The equivalent downstream impedance from the relay:
\begin{align*}
Z_{1d} = & Z_1 || Z_d &
Z_{2d} = & Z_2 || Z_d
\end{align*}

\noindent The positive-sequence current, as a consequence of a line-line fault, flowing through the relay:
\begin{equation*}
I_{ar}^1 = \frac{V_s}{Z_1 + Z_{2d}}
\end{equation*}

\noindent The negative-sequence current flowing through the relay is $I_r^2 \approx I_r^1$ for $Z_{1M}^1 \ll Z_L$. The measured impedance by the relay:
\begin{equation*}
Z_r = \frac{V_{br} - V_{cr}}{I_{br} - I_{cr}} = \frac{V_{ar}^2 - V_{ar}^1}{I_{ar}^1 + I_{ar}^2}
\end{equation*}

\noindent For an upstream relay, where the difference $V_{br} - V_{cr}$ is very small while $I_{br}$ and $I_{cr}$ are not, the measured fault impedance is near zero.

\newpage
\noindent \textit{2) Current-Limiting Inverter and Upstream Relay}

\begin{figure}[!htbp]
\centering
\includegraphics[scale=0.375]{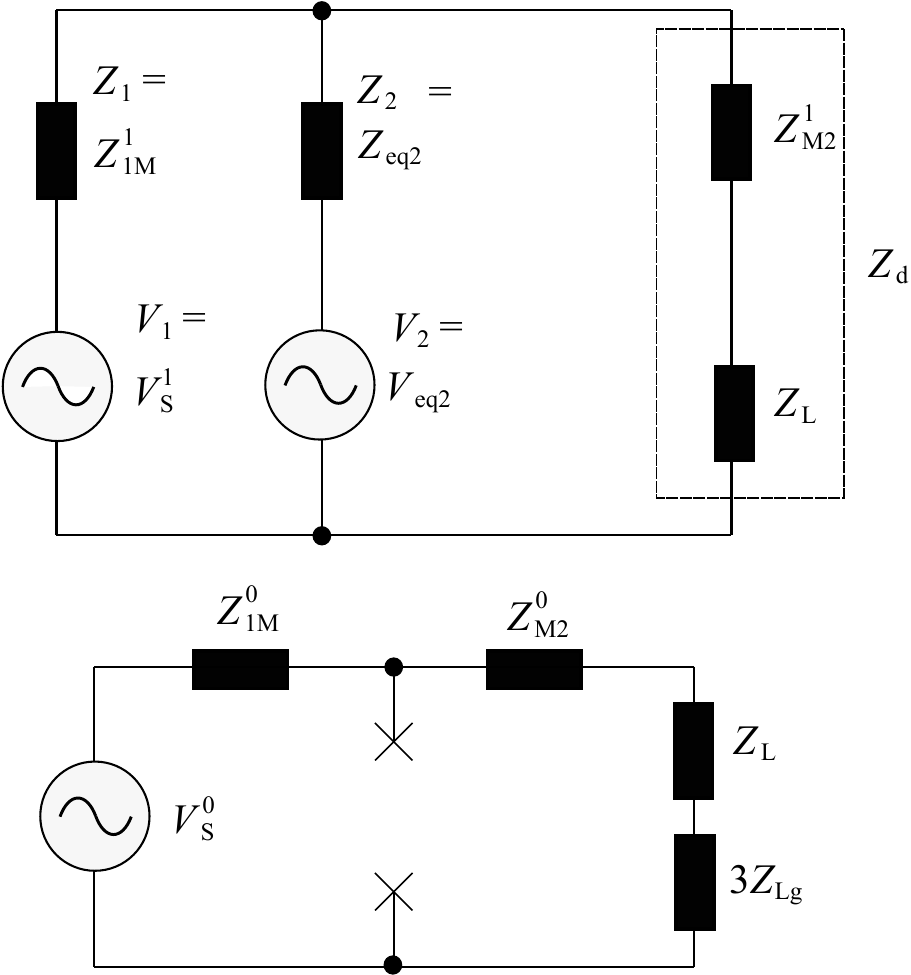}
\caption{Simplified sequence network for a midpoint line-line fault with a current-limiting inverter.}
\label{fig:ug-fault-ll-midpoint-012-simpl}
\end{figure}

In case a current-limiting inverter is used, although the zero-sequence network remains isolated, the inverter now provides zero-sequence voltage; consequently, this network can no longer be neglected in the fault analysis.
To calculate the measured fault impedance, first the interconnection of sequence networks needs to be simplified (illustrated in Fig.~\ref{fig:ug-fault-ll-midpoint-012-simpl}).

First, similarly to Section~III-A-2, the negative-sequence voltage source is disabled. The positive-sequence current, resulting from the positive-sequence voltage source, measured through the relay:
\begin{equation*}
I_{11} = \frac{V_1}{Z_1 + Z_{2d}}
\end{equation*}

\begin{figure}[!htbp]
\centering
\includegraphics[scale=0.375]{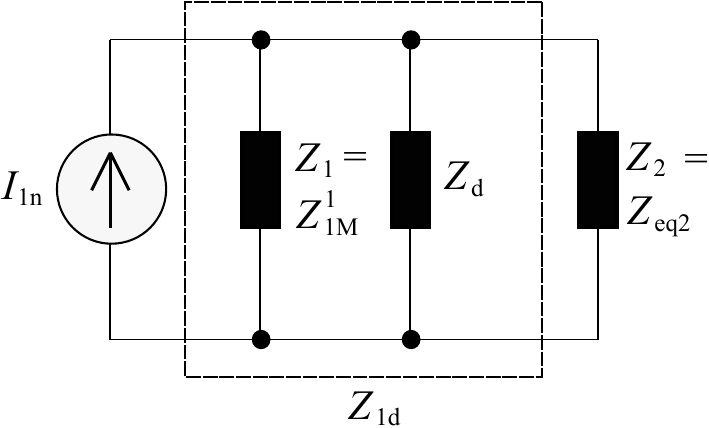}
\caption{Norton equivalent circuit for a line-line fault, supplied by a current-limiting inverter, with the negative-sequence voltage source disabled.}
\label{fig:ug-fault-ll-midpoint-012-norton-v1}
\end{figure}

Second, the Norton equivalent circuit (illustrated in Fig.~\ref{fig:ug-fault-ll-midpoint-012-norton-v1}), with the negative-sequence voltage source disabled, is used.
The negative-sequence current, resulting from the positive-sequence voltage source, flowing through the relay:
\begin{equation*}
 I_{21} \approx I_{1n} \frac{Z_{1d}}{Z_{1d} + Z_2}
\end{equation*}
\noindent assuming that $Z_{1M}^1 \ll Z_L$.

Third, the positive-sequence voltage source is disabled. The negative-sequence current flowing through the relay:
\begin{equation}
I_{22} \approx \frac{V_2}{Z_2 + Z_{1d}}
\end{equation}
\noindent assuming $Z_{1M}^1 \ll Z_L$

\begin{figure}[!htbp]
\centering
\includegraphics[scale=0.375]{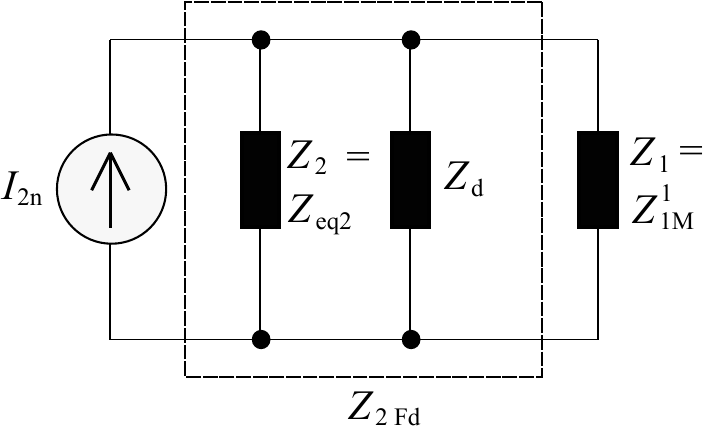}
\caption{Norton equivalent circuit for line-line fault, supplied by a current-limiting inverter, with positive-sequence voltage source disabled.}
\label{fig:ug-fault-ll-midpoint-012-norton-v2}
\end{figure}

Finally, the Norton equivalent circuit (illustrated in Fig.~\ref{fig:ug-fault-ll-midpoint-012-norton-v2}), with the negative-sequence voltage source disabled, is used. The positive-sequence current, from the negative-sequence voltage source, flowing through the relay:
\begin{equation*}
I_{12} \approx I_{2n}
\end{equation*}

The measured impedance can be calculated similarly to Section~III-A-2; as with the ideal voltage source, it is near zero for neglible fault impedance.

\vspace{0.1in}
\noindent \textit{3) Downstream Relay}

By using the definition of sequence voltages \cite{bergen_power_2009} and Fig.~\ref{fig:ug-fault-ll-midpoint-012-bal-simpl}, the measured impedance by the relay:
\begin{align*}
Z_r = & \frac{V_r^b - V_r^c}{I_r^b - I_r^c}=  \frac{(\alpha^2 - \alpha)V_{ar^1} + (\alpha - \alpha^2)V_{ar}^2}{(\alpha^2 - \alpha)I_{ar^1} + (\alpha - \alpha^2)I_{ar^2}} \\
= & \frac{(\alpha^2 +\alpha)(V_{a}r^1 - V_{ar}^2)}{(\alpha^2 + \alpha)(I_{ar}^1 - I_{ar}^2)} = \frac{V_{a}r^1 - V_{ar}^2}{I_{ar}^1 - I_{ar}^2}
\end{align*}

Substituting in the downstream impedances:
\begin{align*}
Z_d^1 = & Z_{M2}^1 + Z_L &
Z_d^0 = & Z_{M2}^1 + Z_L + 3Z_{Lg}
\end{align*}
\noindent the measured impedance can be simplified:
\begin{equation*}
Z_r = \frac{Z_d^1 I_{ar}^1 - Z_d^1 I_{ar}^2}{I_{ar}^1 - I_{ar}^2} =  \frac{Z_d^1(I_{ar}^1 - I_{ar}^2)}{I_{ar}^1 - I_{ar}^2} = Z_d^1
\end{equation*}

This relies on quantities $V_{br}$, $V_{cr}$, $I_{br}$, and $I_{cr}$ being nonzero, which is the case for a practical fault with nonzero fault impedance.

\section{Transient Model} \label{sec:transient-model}
\indent

To validate the above calculated impedances, produced from the equivalent sequence networks, the considered two bus microgrid system is modeled in the MATLAB/Simulink\textsuperscript{\textregistered} SimScape multi-physics 
simulation environment, using the Specialized Power Systems library.
This model is based on the design of an inverter using a PR controller 
\cite{teodorescu_pr_2003} presented in \cite{vasquez_modeling_2013}.

\begin{figure*}[!htbp]
\centering
\includegraphics[scale=0.275,angle=0]{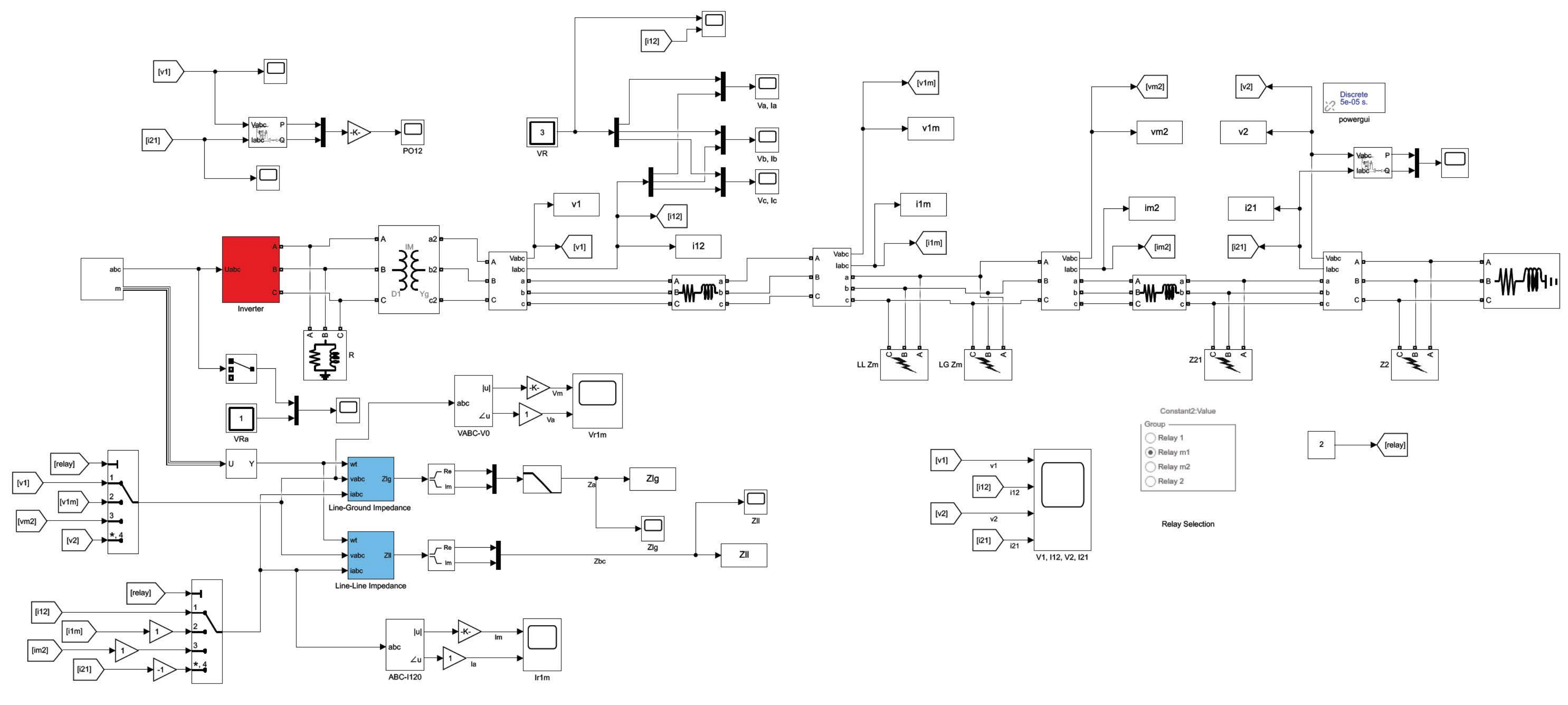}
\caption{Transient microgrid model.}
\label{fig:ug-transient-model}
\end{figure*}

\begin{figure*}[!htbp]
\centering
\includegraphics[scale=0.225,angle=0]{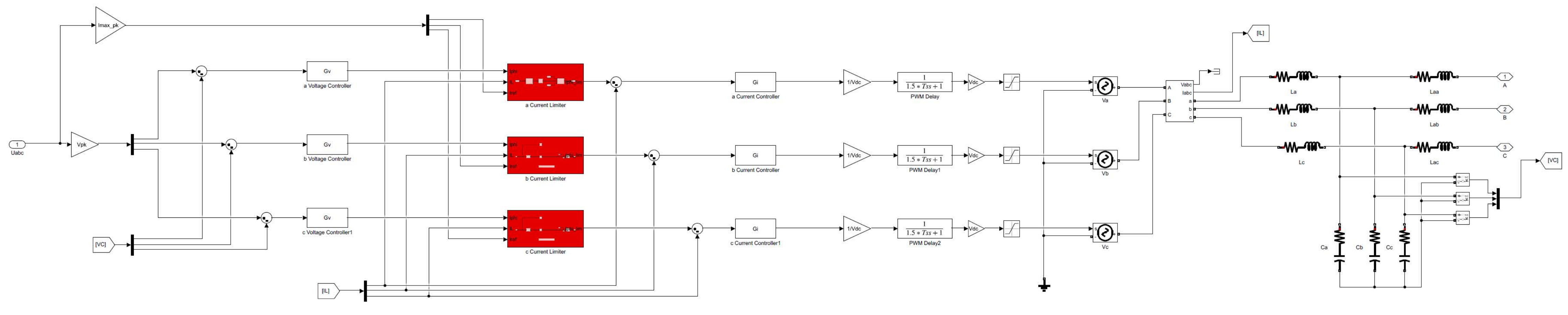}
\caption{Transient inverter model.}
\label{fig:inverter-transient-model}
\end{figure*}

\begin{figure*}[!htbp]
\centering
\includegraphics[scale=0.3,angle=0]{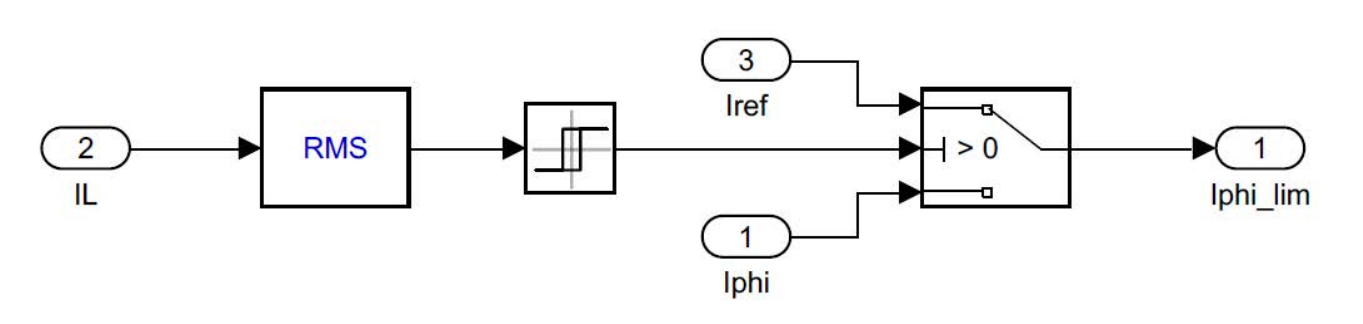}
\caption{Transient current limiter model.}
\label{fig:current-limiter-transient-model}
\end{figure*}

The microgrid is illustrated in Figs.~\ref{fig:ug-transient-model}~--~\ref{fig:current-limiter-transient-model}, while the system parameters are listed in Tables~~\ref{table:global-params}~--~\ref{table:hardware-params}.

\begin{table}[!htbp]
\centering
\caption{Global Parameters}
\begin{tabular}{l l r l} 
 \hline\hline
Name & Symbol & Value & Unit \\ [0.5ex] 
 \hline
Grid frequency & {\tt f0} & 60 & Hz \\
Line-line voltage & {\tt V} & 480 & V \\
 \hline\hline
\end{tabular}
\label{table:global-params}
\end{table}

\begin{table}[!htbp]
\centering
\caption{Synchronous Generator Parameters}
\begin{tabular}{l l r} 
\hline\hline
Subsystem & Symbol & Value \\ [0.5ex] 
\hline
Voltage loop & {\tt kpv} & 0.35 \\
Voltage loop & {\tt krv} & 400 \\
Voltage loop & {\tt kvh5} & 4 \\
Voltage loop & {\tt kvh7} & 20 \\
Voltage loop & {\tt kvh11} & 11 \\
Current loop & {\tt kpi} & 0.7 \\
Current loop & {\tt kri} & 400 \\
Current loop & {\tt kih5} & 30 \\
Current loop & {\tt kih7} & 30 \\
Current loop & {\tt kih11} & 30 \\
\hline\hline
\end{tabular}
\label{table:voltage-loop-params}
\end{table}

\begin{table}[!htbp]
\centering
\caption{Inverter Controller Parameters}
\begin{tabular}{l l r} 
\hline\hline
Subsystem & Symbol & Value \\ [0.5ex] 
\hline
Voltage loop & {\tt kpv} & 0.35 \\
Voltage loop & {\tt krv} & 400 \\
Voltage loop & {\tt kvh5} & 4 \\
Voltage loop & {\tt kvh7} & 20 \\
Voltage loop & {\tt kvh11} & 11 \\
Current loop & {\tt kpi} & 0.7 \\
Current loop & {\tt kri} & 400 \\
Current loop & {\tt kih5} & 30 \\
Current loop & {\tt kih7} & 30 \\
Current loop & {\tt kih11} & 30 \\
\hline\hline
\end{tabular}
\label{table:inverter-controller-params}
\end{table}

\begin{table}[!htbp]
\centering
\caption{Hardware Parameters}
\begin{tabular}{l l r l} 
 \hline\hline
Name & Symbol & Value & Unit \\ [0.5ex] 
 \hline
Inverter rated power & {\tt P} & 50 & kW \\
DC-bus voltage & {\tt Vdc} & 1800 & V \\
Output filter inductance & {\tt L} & 18 & $\mu$F \\
Output filter capacitance & {\tt C} & 250 & nF \\
Maximum rms output current & {\tt Imax} & 70 & A \\
Cable resistance & {\tt Rc} & 39 & m$\Omega$ \\
Cable inductance & {\tt Lc} & 70.8 & $\mu$H \\
Load real power & {\tt Pd} & 25 & kW \\
Load reactive power & {\tt Qd} & 12.5 & kW \\
 \hline\hline
\end{tabular}
\label{table:hardware-params}
\end{table}

\section{Results} \label{sec:results}
\indent

\begin{figure*}[!htbp]
\centering

\subfloat[Line-ground fault with an upstream ground distance relay]{\includegraphics[width=2.6in]{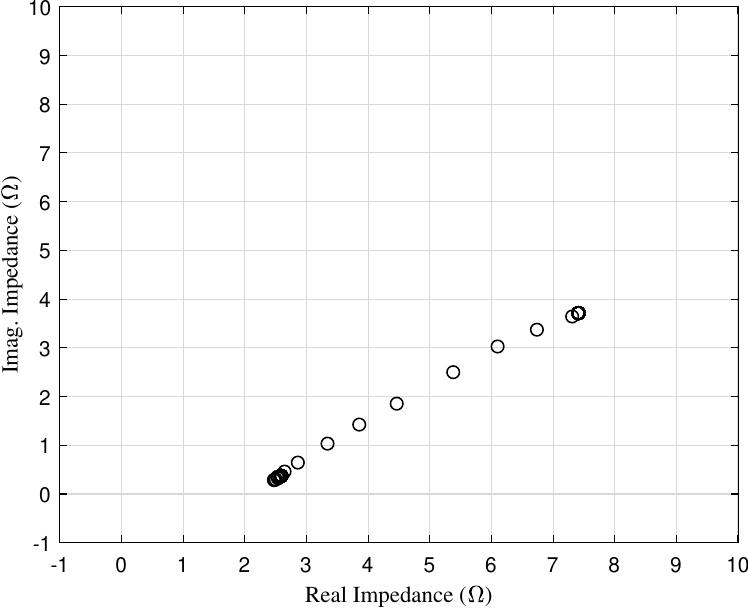}
\label{fig:mp-lg-fault-zlg-upstream}}
\hfil
\subfloat[Line-ground fault with a downstream ground distance relay]{\includegraphics[width=2.65in]{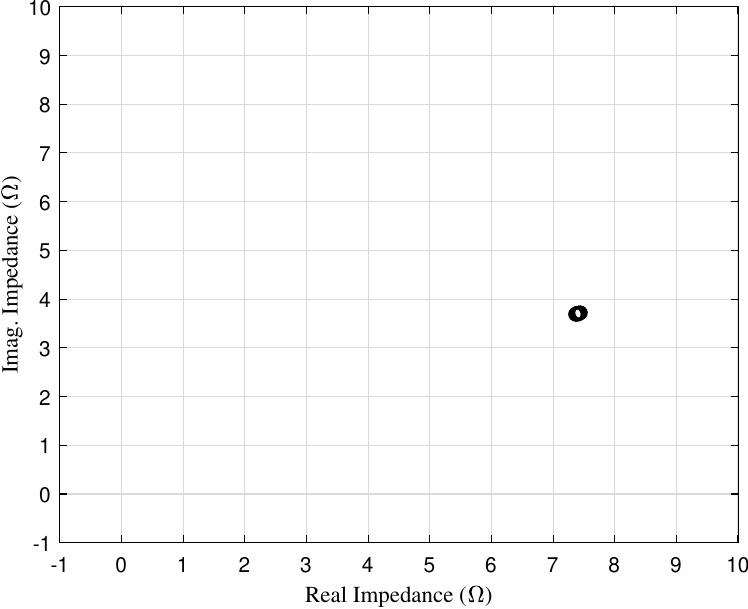}
\label{fig:mp-lg-fault-zlg-downstream}}

\subfloat[Line-line fault with an upstream phase distance relay]{\includegraphics[width=2.6in]{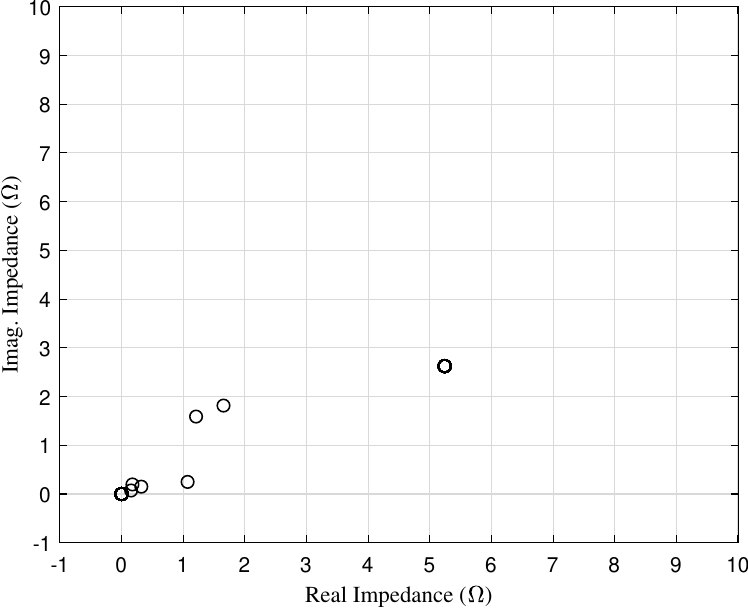}
\label{fig:mp-ll-fault-zll-upstream}}
\hfil
\subfloat[Line-line fault with a downstream phase distance relay]{\includegraphics[width=2.6in]{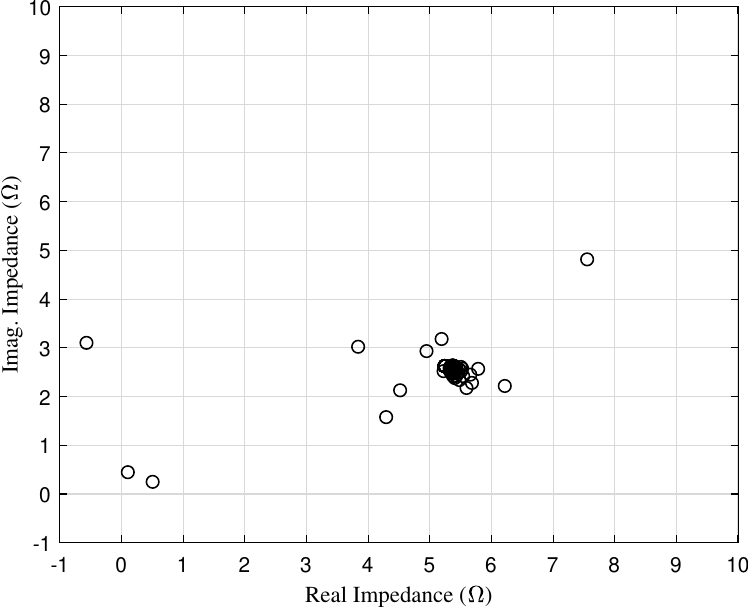}
\label{fig:mp-ll-fault-zll-downstream}}

\vspace{0.1in}
\caption{Measured impedances on different faults with different distance relays for the four cases considered.}
\label{fig:case-results}
\end{figure*}

Fig.~\ref{fig:case-results} presents the simulation results for the four cases considered: (a) line-ground fault with an upstream ground distance relay, (b) line-ground fault with a downstream ground distance relay, (c) line-line fault with an upstream phase distance relay, and (d) line-line fault with a downstream phase distance relay.
For these cases, only the current-limiting inverter is considered and the ideal voltage source case is not modeled.

Figs.~\ref{fig:mp-lg-fault-zlg-upstream} and \ref{fig:mp-ll-fault-zll-upstream} illustrate that for both upstream relay cases, the measured impedance varies over time from the load resistance towards the origin, while remaining in the first quadrant.
Figs.~\ref{fig:zr1-ug-fault-lg-midpoint} and \ref{fig:mp-lg-fault-zlg-upstream} highlight the correspondence between the predicted and measured impedances.
Fig.~\ref{fig:mp-ll-fault-zll-upstream} confirms that the measured impedance for a relay upstream of a line-line fault is near zero.
Figs.~\ref{fig:mp-lg-fault-zlg-downstream} and \ref{fig:mp-ll-fault-zll-downstream} proves that the measured impedance for downstream relays is approximately the load impedance in both cases.

\newpage
\section{Conclusions} \label{sec:conclusions}
\indent

Both the analytical and simulation results suggest that conventional relaying quantities for phase and ground distance protection are suitable for the protection of both lines and load buses in inverter-interfaced microgrids.
Although the case studies of this paper used midpoint faults, the measured quantities change only a small amount for load bus faults as $Z_{1M}^1 \ll Z_L$.
For the case of line protection, it is likely that pilot protection is necessary rather than relying on zone-based distance protection. The low line impedances of microgrids make it likely that other impedances (e.g., inverter virtual impedance, load impedance, fault impedance) dominate, rendering it difficult to distinguish in- or out-of-zone faults based purely on measured impedance.

The results do not support the observations of Dewadasa et al.~\cite{dewadasa_line_2008, dewadasa_distance_2008} on protection misperations for an upstream line-ground fault when a grounded-wye load is present.
While the performed case studies were not able to determine why Dewadasa observed misoperations, from Fig.~\ref{fig:ug-fault-lg-midpoint-seqs}c it is apparent that an inverter lacking a grounding source, causing $Z_{1M}^0$ to be open-circuited, could result in downstream ground overcurrent protection without directional relaying to trip. Inspection of Fig.\ref{fig:ug-fault-lg-midpoint-seqs}b confirms the claim of \cite{dewadasa_line_2008} that negative-sequence current is suitable for use as a polarizing quantity for directional relaying in line-ground faults.


\vspace{0.1in}
\bibliographystyle{unsrt}
\bibliography{references}

\end{document}